\documentclass[12pt,manuscript]{aastex}

\usepackage{amsmath}

\shorttitle{Radiation hardness studies of InGaAs and Si photodiodes at 30, 52, \& 98 MeV}
\shortauthors{Baptista et al.}

\begin{document}

\title{Radiation hardness studies of InGaAs and Si photodiodes at \\ 30, 52, \& 98 MeV and fluences to  $ 5\times10^{11}$ protons/cm$^2$}
\author{Brian J. Baptista and Stuart L. Mufson}
\affil{Department of Astronomy, Indiana University }
\affil{727 E. Third St. Bloomington, IN 47405}
\email{bbapti@astro.indiana.edu}

\begin{abstract}
Here we report the results of an investigation into the effects of ionizing radiation  on commercial-off-the-shelf InGaAs and Si photodiodes.  The photodiodes were exposed to 30, 52, and 98 MeV protons with fluences ranging from  $10^{8} - 5\times10^{11}$ protons/cm$^2$ at the Indiana University Cyclotron Facility.  We tested the photodiodes for changes to their dark current and their relative responsivity as a function of wavelength.  The Si photodiodes showed increasing damage to their responsivity with increasing fluence; the InGaAs photodiodes showed significantly increased dark current as the fluence increased.  In addition, we monitored the absolute responsivity of the InGaAs photodiodes over their entire bandpass.  Our measurements showed no evidence for broadband degradation or graying of the response at the fluences tested.  All measurements in this investigation were made relative to detectors traceable to NIST standards.  
\end{abstract}


\section{Introduction}
The accurate calibration of optical and near infrared photometry was a primary objective of SNAP/JDEM \citep{bib:SNAP1,bib:SNAP2}, since renamed WFIRST \citep{bib:WFIRST}, a space-based mission to study dark energy.  In particular, systematics-limited measurements of the dark energy parameters based on SNIa observations required broadband filter measurements with a top-down error budget of 2\% color error and a 1.5\% in-band error \citep{bib:nickDissertation}.  To achieve these ambitious goals SNAP/JDEM considered a multi-technique approach to focal plane calibration that included accurate stellar photometry and an onboard flat fielding illumination system.  For the low-frequency spatial flats (L-flats) that characterize large scale, many-pixel variations across the focal plane, the technique favored was to dither large ensembles of stars across the focal plane and to measure the stars' offsets  from the ensemble mean \citep{bib:vanderMarel,bib:nickDissertation}.  On the other hand, high-frequency spatial flats (S-flats), or small scale few-pixel variations across the focal plane, are typically monitored by a flat-fielding illumination system.  For SNAP/JDEM, one version of this onboard calibration light system included filament lamps and/or pulsed LEDs, light sources that were to be driven at constant current.  Since the accuracy of such high-frequency flat fielding systems depends strongly on the capacity to compensate for variations in the illuminators, stable and precisely calibrated photodiode detectors were included to monitor the light sources.  

Modern calibration strategies to achieve flat fielding goals as challenging as those required for the SNAP/JDEM high frequency flats are typically based upon monitoring photodiodes calibrated to the state-of-the-art by NIST or equivalent national standards laboratories.  Since SNAP/JDEM was planned as a multiyear mission at the L2 Lagrange point, approximately $1.5 \times 10^6$ km from Earth, we studied these monitoring photodiodes for their ability to survive in the radiation environment expected.  As described by \citet{bib:dawson-bebek-snap-radenvrioment}, solar protons dominate the radiation environment at L2.  For instance, in a six year mission with an experiment carrying a minimum Aluminum equivalent shielding of 9.0 mm with an average shielding thickness of 38 mm, the focal plane monitoring photodiodes would be exposed to fluences of order ${\mathcal F} = 10^9 - 10^{10}$~cm$^{-2}$ in the range $10 - 100$ MeV.
 
In this investigation, we describe radiation hardness studies of several commercial off-the-shelf (COTS) InGaAs photodiodes and one Si photodiode irradiated with protons at energies 30, 52, and 98 MeV, and with fluences up to  ${\mathcal F} =  5 \times 10^{11}$~cm$^{-2}$ at the  Indiana University Cyclotron Facility (IUCF).   In our detector-based calibration scheme for SNAP/JDEM, we proposed InGaAs photodiodes to monitor the NIR flat-field illuminators and Si photodiodes to monitor the optical illuminators.  The intent was to have all the photodiodes  calibrated by NIST.  Here we examine both the relative spectral response and the overall broadband response of InGaAs and Si photodiodes as a function of the irradiating proton energy and fluence.  Although there have been radiation hardness studies of Si photodiodes, including CCDs \citep{bib:dawson-bebek-snap-radenvrioment}, there is surprisingly little information on InGaAs photodiodes.  As proposed space missions push into the NIR (EUCLID, WFIRST), InGaAs photodiodes are likely to find wider use in space-based experiments. 

\section{Photodiodes and Radiation Effects Research Program}
\label{sec:experimentalProgram}
In this section we describe the photodiodes tested and the facility where the radiation exposures were made.

\subsection{Photodiodes}
The photodiodes tested in these studies were obtained from five optoelectronic manufacturers.  All were COTS devices and are listed in Table~\ref{tab:Photodiode}.  The InGaAs photodiodes all have a responsivity cutoff at 1700~nm and the Si photodiodes have the typical responsivity cutoff at 1000~nm.  When possible, the photodiodes were obtained in TO-5 packaging, which has been shown to be robust to the shake and heat cycle testing required to qualify for space flight\footnote{http://www.osioptoelectronics.no/custom-oem-solutions/Space.asp}.  Only the InGaAs photodiodes from Advanced Photonics were unavailable with this packaging as a COTS device.  Since our plan was to expose the InGaAs photodiodes to protons at three energies, we obtained three InGaAs photodiodes from each manufacturer.  We also obtained an additional PerkinElmer InGaAs photodiode that was not exposed to radiation and which was used as a control to determine systematic errors.  
We used 7 OSI Si photodiodes  in this study, one for each fluence we planned on using.  We only needed 7 Si photodiodes because they were irradiated at only one energy.   

The spectral response measurements of the test photodiodes were made with respect to stable reference photodiodes whose response was calibrated by NIST.  The NIST-calibrated reference InGaAs and Si photodiodes are  listed in Table~\ref{tab:Photodiodes-nist}.

\subsection{ Radiation Effects Research Program }
The photodiodes were irradiated at the Radiation Effects Research Program (RERP) facility at IUCF\footnote{http://www.iucf.indiana.edu/rerp/}.  The IUCF cyclotron produces 200 MeV protons with dosimetry better than 10\% \citep{bib:RERP-beamprop}.  The energies chosen for the tests, 30, 52 and 98 MeV, were selected to span a reasonable range of the particle environment expected at L2 \citep{bib:dawson-bebek-snap-radenvrioment} from the range of energies available at RERP.  Energies lower than 30 MeV are not available.  RERP has both wide band and narrowband beam configurations.  The wide band beam degrades the energy of the 200 MeV protons to the program energy using a thick copper plate upstream of the target photodiodes.  In the narrowband beam configuration, the protons first pass through a magnetic spectrometer that selects out a narrow range of momenta and then these selected protons pass through a beryllium degrader to obtain the program energy.  The advantage of the narrowband beam is that the radiation hardness studies can focus on specific damage mechanisms.  The advantage of the wide band beam is that the proton flux is two orders of magnitude greater than the narrowband beam, thereby reducing exposure times.  At 45 MeV the energy resolution in the narrow  band beam is approximately 3.3~MeV~(FWHM); for the broadband beam, the energy resolution is approximately 13.5~MeV~(FWHM).  

For fluences ${\mathcal F} \leq 5\times10^9$ cm$^{-2}$, we used the narrowband beam and exposure times were a few minutes long.  For fluences ${\mathcal F} > 5\times10^9$ cm$^{-2}$, we used the wide band beam to keep the exposure times reasonably short.  Exposure information is given in Table~\ref{tab:radruns}.  During radiation exposure, all of the target photodiode's pins were connected and grounded.

\section{Dark Current and Relative Spectral Response}

These investigations were primarily aimed at understanding whether the spectral response of photodiodes degrades in response to the ionizing radiation environment expected at L2.  

\subsection{Indiana Relative Responsivity Measurement Apparatus}
\label{sec:IRRMA}

\begin{figure}
\plotone{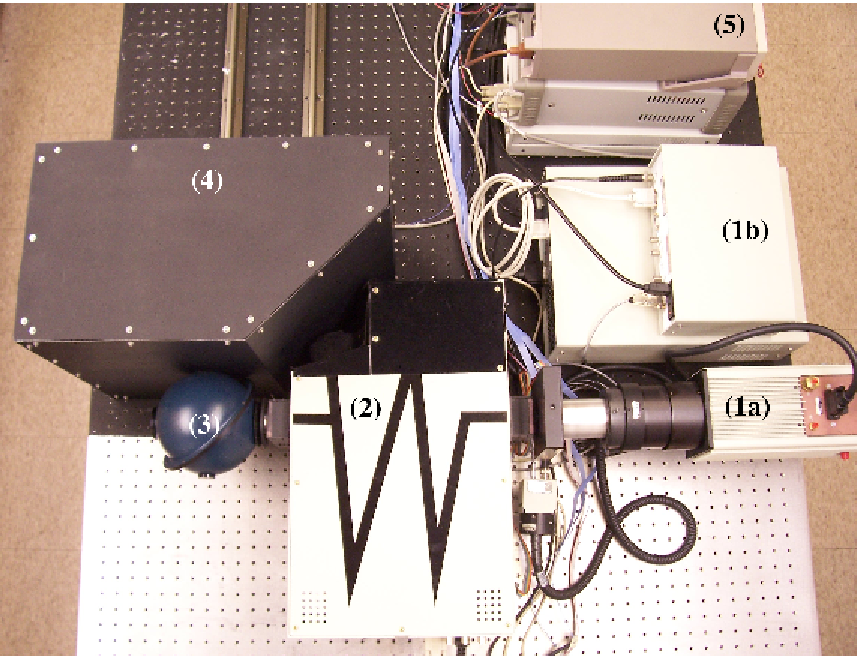}
\caption{The IRRMA apparatus for the measurement of dark current and relative spectral response.}
\label{fig:hardware}
\end{figure}

We developed the Indiana Relative Responsivity Measurement Apparatus (IRRMA) to measure dark current and relative spectral response as a function of wavelength; the apparatus is shown in Figure \ref{fig:hardware}.  The apparatus consists of a radiance-controlled QTH lamp source powered by a constant-current power supply, a monochromator that feeds into an integrating sphere, and a dual-channel picoammeter that measures the output current of the test photodiode and a NIST-calibrated reference photodiode simultaneously.  The NIST-calibrated photodiode is temperature controlled.  The light source (1a) is a 100 W incandescent QTH lamp in a Newport Photomax housing.  The lamp is powered by a radiometric power supply (1b) controlled by a Newport digital exposure controller.  The Newport monochromator (2) has two gratings that span the wavelength range investigated here: 500--1000~nm for Si photodiodes and 1000--1600~nm for InGaAs photodiodes.  The monochromator slits produce a 4 nm bandpass at the output.  There is a shutter at the input of the monochromator that provides a measurement of the dark current when closed.  The output of the 4$''$ Labsphere integrating sphere (3) projects flat illumination onto both the test photodiode and the NIST-calibrated reference photodiode.  The two photodiodes are mounted in the dark box (4) where they simultaneously view the light from the integrating sphere.  A Keithley dual-channel picoammeter (5) measures the two photodiodes with zero external bias.  However, to make the current measurement, the picoammeter by design applies a measured $60\pm5 \mu V$ reverse bias to the photodiode which in turn generates a few picoamps of dark current.    
 
The picoammeter sampled the output of the photodiodes at 6 Hz at every 2 nm wavelength step.  To reduce read noise, 30 measurements were taken at each wavelength step and averaged.  The dark current was measured every 10th wavelength step, or at 20 nm intervals.  The dark current was subtracted from the signal at each 2 nm step by computing a linear interpolation of the dark current over each 20 nm measurement interval.   For the results described in \S\ref{results} we made two full wavelength scans  of each photodiode after an exposure at IUCF and we averaged the dark subtracted measurements.  The dark current results we report below are the mean of the dark current measurements for both scans. 

\subsection{Experiment Design}

After each radiation exposure, we measured the spectral response of the test photodiodes in IRRMA concurrently with respect to the stable NIST-calibrated reference photodiode.  The reference photodiodes were calibrated by NIST with standard spectroradiometric detector calibration services\footnote{http://www.nist.gov/calibrations/spectroradiometric.cfm\#39075S}.  After each radiation exposure we made measurements of the test photodiode at each wavelength step and compared these to the reference photodiode at the same wavelength.  To minimize systematic errors, we used the fractional change in the ratio of the responses of the test photodiode to the reference photodiode as the test statistic.

The complication with reducing systematic errors in this way, however, is that differences between the response of the test and reference photodiodes can mask or exaggerate real spectral changes due to radiation exposure.  If the spectral response of the reference photodiode at a particular wavelength is large compared with the test photodiode, for instance, significant spectral changes in the test diode would not change the ratio appreciably, thereby masking real spectral changes.  On the other hand, if the spectral response of the reference diode is small compared with the test diode, small spectral changes in the test diode would lead to large changes in the ratio.  To account for different spectral responses of the test and reference photodiodes, we normalized the ratios with respect to those measured at $\lambda_{norm}$ = 600~nm for Si photodiodes or at $\lambda_{norm}$ = 1200~nm for InGaAs photodiodes.  By normalizing the ratio in this way, we clearly only measured {\it relative} changes in the spectral response of the test photodiodes.  

To be quantitative, we defined $\Delta R_\lambda(t)$ as our test statistic. 
Let $I_\lambda(t)$ = photocurrent response of the test photodiode to the QTH lamp at wavelength $\lambda$ measured in the $\Delta \lambda = 4$~nm monochromator bandpass at some time $t$ during the experiment and let $D(t)$ = the interpolated dark current at time $t$.  Similarly, let $N_\lambda$ = photocurrent response of the NIST-calibrated reference photodiode to the QTH lamp at wavelength $\lambda$ and $d$ = its interpolated dark current during the scan.  Aside from variations in the lamp output, $N_\lambda$ and $d$ are assumed to have no additional time-dependent behavior.  
Then the normalized response of the test photodiode relative to the NIST calibrated photodiode at wavelength $\lambda$ at time $t$, $R_\lambda(t)$, is given by  
\begin{equation}
\label{eqn:NTNR}
 R_\lambda(t) = \left[\frac{I_\lambda(t) -D(t)}{N_\lambda- d}\right] \left/\left[\frac{I_{\lambda}(t) - D(t)}{N_{\lambda}- d}\right]_{\lambda = \lambda_{norm}}\right . ,
\end{equation}
where only the bracketed expression in the denominator is evaluated at the wavelength $\lambda_{norm}$.
We quantified changes in photodiode response  as a result of radiation exposure with the fractional change in $R_\lambda(t)$,
\begin{equation}
\label{eqn:deltaNTNR}
\Delta R_\lambda(t) = R_\lambda(t)/R_\lambda(0) - 1.
\end{equation}

\subsection{Systematic Uncertainties}

\begin{figure}
\plotone{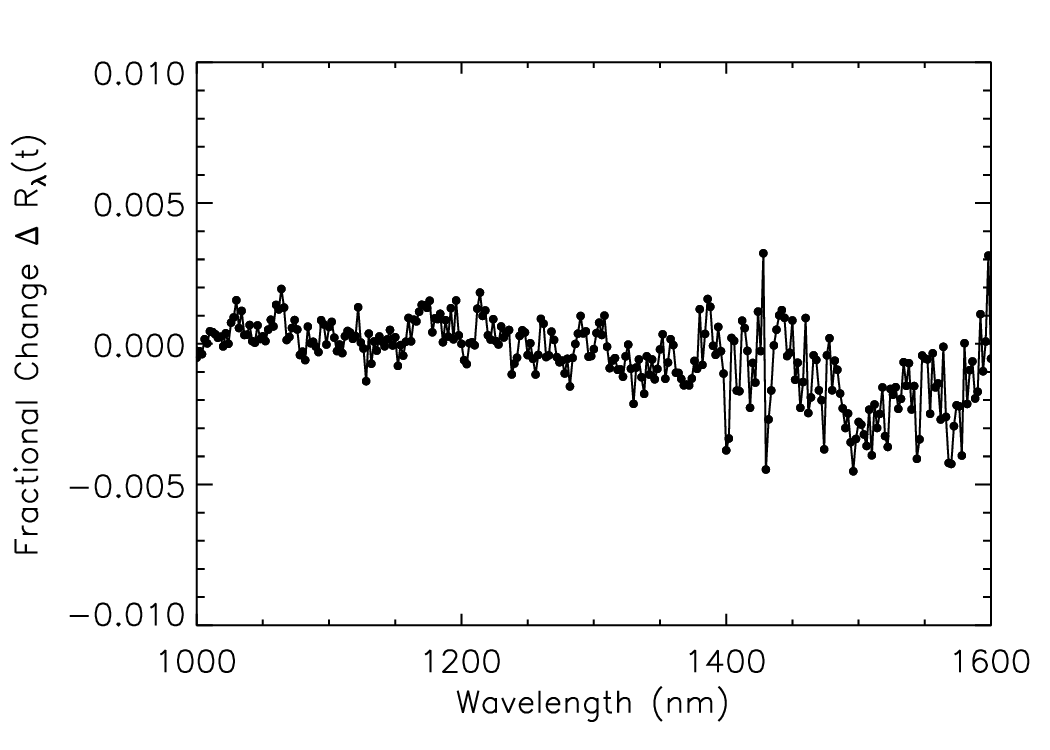} 
\caption[]{Variations in the response of a Perkin-Elmer InGaAS photodiode introduced by a 4.1$^\circ$C temperature variation from 13.4$^\circ$C to 17.5$^\circ$C.  The systematic variations are less than 0.5\%.}
\label{fig:tempStability}
\end{figure}

Several systematic effects in the apparatus that could introduce uncertainties that mask real changes in the spectral response of the test photodiodes cancel in the ratio defining $R_\lambda(t)$ in eq.(\ref{eqn:NTNR}).  For instance, QTH lamps have rated lifetimes of 50 hr and needed to be replaced on occasion during the course of these investigations.  Since both the test photodiode and the NIST reference photodiode see the same lamp light, these systematics cancel in $R_\lambda(t)$.  One systematic effect not canceled in this way are variations in the test photodiode responsivity introduced by measurements in our laboratory, which is not equipped to maintain a constant temperature for the test photodiode.  Since the responsivity of photodiodes as a function of temperature can differ depending on the manufacturer, this systematic effect would not be canceled in $R_\lambda(t)$.
We tested for this systematic effect by measuring  $\Delta R_\lambda$ for an unexposed control PerkinElmer InGaAs photodiode at 13.4$^\circ$C and 17.5$^\circ$C, a temperature range greater than we encountered during our measurements.  In these measurements, we attached a TEC to the back of the photodiode package and used an Omega controller to vary the temperature.  The results are shown in Fig.~\ref{fig:tempStability}, which shows clearly that negligible systematic variations of $< 0.5\%$ are introduced by temperature variations over this range.

Before radiation exposure, we measured the dark current for all of the test photodiodes.  The results of the measurements for the InGaAs photodiodes are given in Table~\ref{tab:baselinedark}; for the Si photodiodes, the dark currents are given in Table~\ref{tab:baselinedarksi}.  Except for the Fermionics photodiodes, the baseline dark currents vary by up to a factor of $\sim$50\%; the Fermionics photodiodes show considerably greater variation.  
Again excluding Fermionics photodiodes, the dark currents vary by a factor of $\sim$2 from manufacturer to manufacturer; the Fermionics photodiodes are by far the noisiest.    

\begin{figure}
\plotone{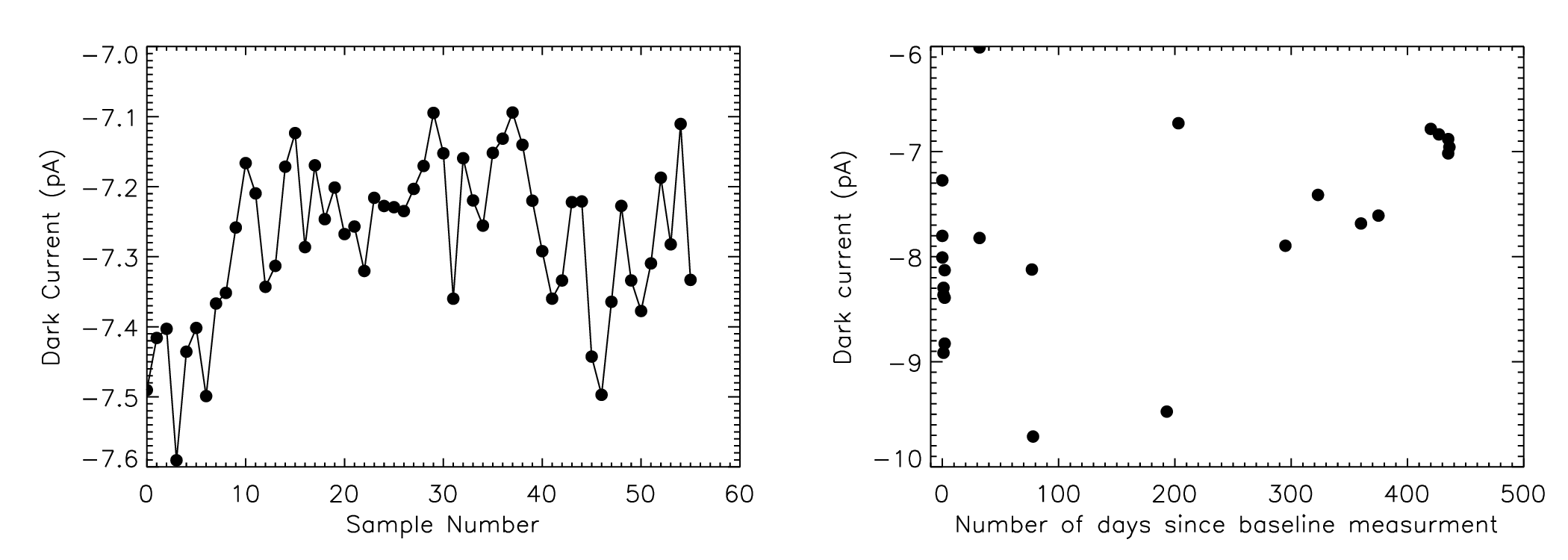}\\
\caption[]{Variations in the dark current for the unexposed control PerkinElmer InGaAs photodiode.  {\it left panel}: variations in the dark current for a typical single full-wavelength scan.  The RMS/mean for this scan is $\sim$2\%.  {\it right panel}: variations in the dark current seen over the $\sim$600 days of the experiment.  For measurements made at the same temperature, the RMS/mean is $\sim$12\%.  }
\label{fig:darkstability}
\end{figure}

We determined the systematic uncertainties in measurements of the dark current by repeatedly sampling the dark current of the control PerkinElmer InGaAs photodiode and assuming the dark current of this photodiode did not change.  
These measurements are shown in Figure~\ref{fig:darkstability}.  
In the left panel, the figure shows typical variations in the dark current during a single full-wavelength scan that takes approximately one hour.  The RMS/mean for these measurements is $\sim$2\%.  
The right panel shows measurements of the dark current over the duration of the experiment ($\sim$600 days) that only includes measurements of the control PerkinElmer photodiode made at the same temperature. 
For these measurements, the RMS/mean in the dark current is $\sim$12\%.
Adding the short term and long term variations in quadrature gives the systematic error in the dark current measurements of $\sim$12\%.

\subsection{Changes in Relative Spectral Response due to Radiation Exposure}

After each exposure, we used IRRMA to test our photodiodes for changes in relative spectral response.  We considered three possibilities.
First, it is possible that radiation exposure did not affect the relative spectral response.  In that case, $\Delta R_\lambda(t)$ will be consistent with zero within measurement errors.  
However, if the relative spectral response did change, then
eq.(\ref{eqn:NTNR}) shows that there are two different ways that radiation could affect the photodiode.  
In one, radiation exposure damages the spectral response $I_\lambda$ of the photodiode, resulting in a wavelength-dependent change in $\Delta R_\lambda$.  In the other, radiation exposure drives an increase in the dark current $D$, resulting in a wavelength-independent change in $\Delta R_\lambda$.   We have developed a simple $\chi^2$ statistic to differentiate among these three possibilities,  
\begin{equation}
\label{eqn:chisq}
\chi^2/NDF = \frac{1}{(n-1)}\sum_{\lambda = \lambda_0}^{\lambda_{f}} \frac{[\Delta R_\lambda(t) - \mathcal{R}_\lambda]^2}{\sigma^2} ,
\end{equation}
where the sum is over the $n$ discrete monochromator measurements in the wavelength range ($\lambda_0, \lambda_{f}$), $\mathcal{R}_\lambda$ is the model for the radiation damage, and $\sigma$ is the RMS error in the measurement of $\Delta R_\lambda$ in our apparatus.

\begin{figure}
\plotone{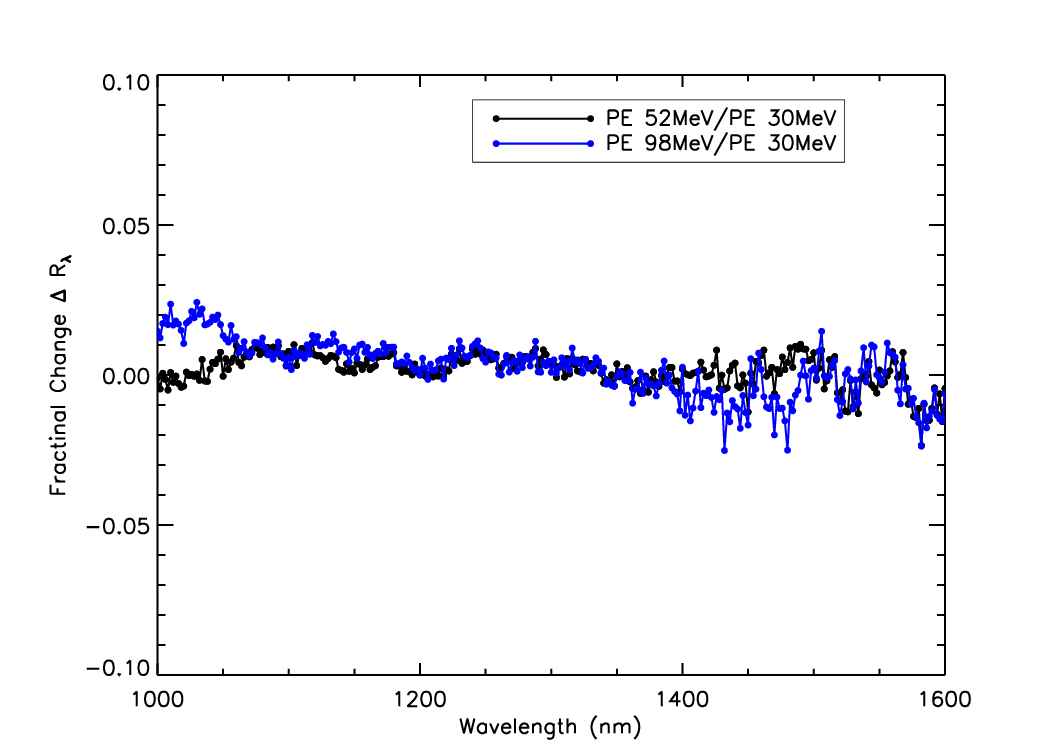}
\caption{The $\Delta R_\lambda$ distributions for the PerkinElmer photodiodes before radiation exposure.  Both the 52 MeV and 98 MeV photodiodes were compared to the 30 MeV photodiode.  The RMS of these distributions was used to determine $\sigma$ in Table~\ref{tab:chisqvalues}.}
\label{fig:sigma}
\end{figure}

Central to this analysis is the determination of $\sigma$, since it sets the scale that differentiates between real physical effects and measurement error.  We determined $\sigma$ for each photodiode individually from the initial $\Delta R_\lambda(0)$ measurements taken prior to radiation exposure.  Since we have three instances of each photodiode, one for each proton energy, we constructed two $\Delta R_\lambda$ data sets -- one in which the 52 MeV data were compared to the 30 MeV data in eq.(\ref{eqn:deltaNTNR}) and one in which the 98 MeV data were compared to the 30 MeV data.  Assuming the three instances of the the photodiodes are identical, we used the RMS of these two distributions to estimate of the $\sigma$ in eq.(\ref{eqn:chisq}).  Fig.~\ref{fig:sigma} shows the distributions of ``identical'' photodiodes for those manufactured by PerkinElmer.  The value of $\sigma$ for the PerkinElmer photodiode, as well as those for the others in Table~\ref{tab:Photodiode}, are given in Table \ref{tab:chisqvalues}.

We then used these $\sigma$'s to test against the three simple models for radiation damage with eq.(\ref{eqn:chisq}).  As usual, our strategy looks for $\chi^2/NDF \sim 1$ for a successful model fit.  
For a photodiode that does not change, $\mathcal{R}_\lambda = 0$.  
If $\chi^2/ndf$ is significantly greater than 1 for the $\mathcal{R}_\lambda = 0$ model, we assumed there was a change in the photodiode response and then tested the data against a model in which $\Delta R_\lambda(t)$ changes with $\lambda$.  For this model we set $\mathcal{R}_\lambda$ equal to a fifth order polynomial fit to the $\Delta R_\lambda(t)$ data at each fluence.  If  $\chi^2/NDF \sim 1$ for this model at each fluence --  that is, the $\Delta R_\lambda(t)$ data can be well described by a wavelength-dependent model within the measurement errors -- we concluded that radiation exposure damages the spectral response $I_\lambda$ of the photodiode.
If  $\chi^2$/NDF is significantly greater than 1 for this model, then we concluded that the changes in $\Delta R_\lambda(t)$ are due to wavelength-independent effects, like an increase dark current $D$ as the fluence increases.  In this case, we looked for a correlation between increases in $\chi^2/ndf$ and dark current $D(t)$.  

\begin{figure}
\plotone{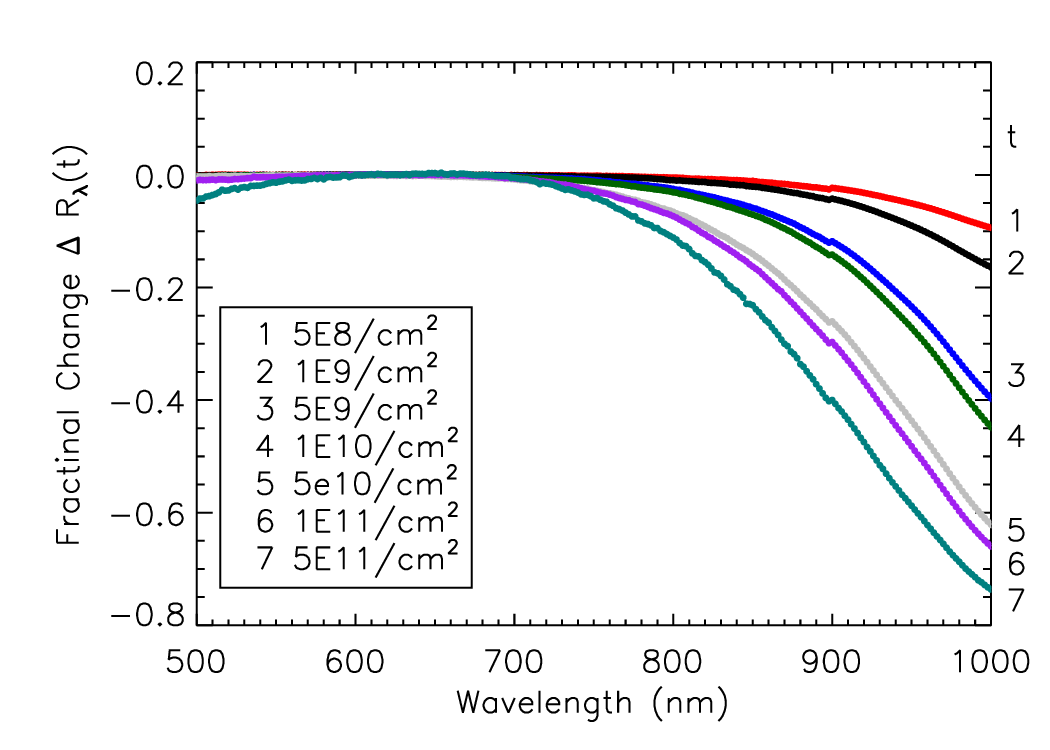}
\caption[]{$\Delta R_\lambda(t)$ for the OSI Si photodiode irradiated at 30 MeV, plotted at seven fluences between $5\times10^8$ and $5\times10^{11}$ protons/cm$^2$.}
\label{fig:OSIsifracdamage}
\end{figure}

\begin{figure}
\plotone{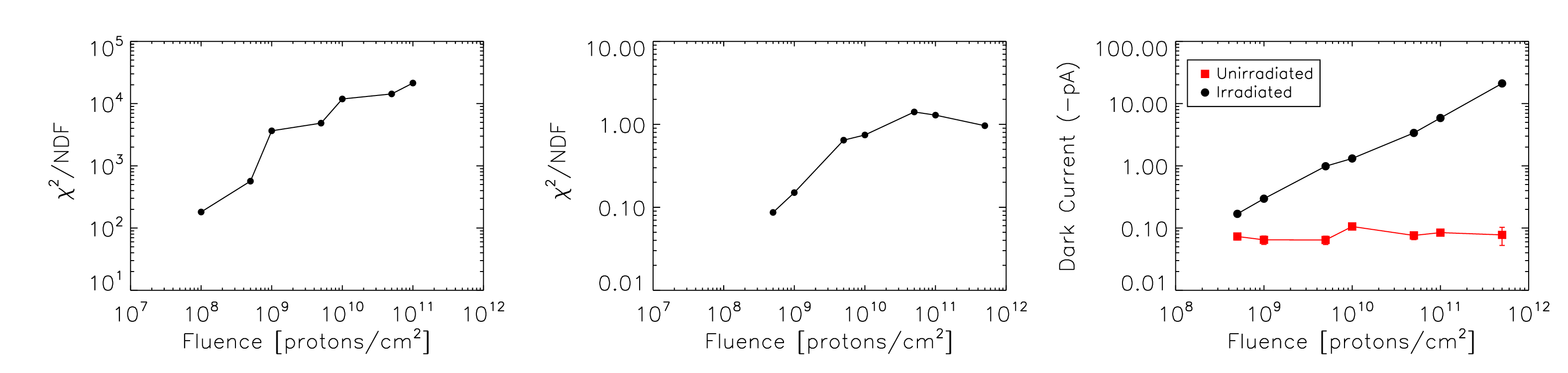}
\caption[]{Model fits to the radiation exposure data for the OSI Si photodiodes.  {\it left}: $\chi^2$/NDF for the $\mathcal{R}_\lambda = 0$ model as a function of fluence for 30 MeV proton irradiation.  Since $\chi^2$/NDF $\gg 1$ this model fits the data poorly, implying radiation damage.  {\it middle}:  $\chi^2$/NDF for the wavelength-dependent model of radiation damage.   Since $\chi^2$/NDF $\lesssim 1$, we conclude that radiation exposure damages the spectral response of the Si photodiode.  {\it right}:  The (negative) dark current as a function of fluence.  Changes are well in excess of our $\sim12$\% measurement uncertainties.  Since the photocurrents dominate the dark currents at all fluences, the dark currents do not strongly affect our measurements of relative responsivity. }
\label{fig:OSIsidarkandchisq}
\end{figure}

We demonstrate our analysis approach with an example.  Fig.~\ref{fig:OSIsifracdamage} shows $\Delta R_\lambda(t)$ for the OSI Si photodiodes irradiated at 30 MeV and fluences between $5\times10^8$ -- $5\times10^{11}$ protons/cm$^2$.  There are clear and significant spectral response changes evident that increase with exposure.  
The left panel of Fig~\ref{fig:OSIsidarkandchisq} shows $\chi^2$/NDF for the $\mathcal{R}_\lambda = 0$ model as a function of fluence.  As expected, this model fit is not a good one.  The middle panel shows $\chi^2$/NDF for the wavelength-dependent model of radiation damage.  Since $\chi^2/NDF \lesssim 1$ for this model, it is clear that radiation exposure damages the spectral response $I_\lambda$ of the Si photodiodes, with increasing damage as the fluence increases.  The Si photodiodes started off with less than  $\sim0.1$ pA of dark current, as shown in Table \ref{tab:baselinedarksi}.  The right panel shows the variations in dark current with radiation exposure.  Changes are well in excess of our $\sim12$\% measurement uncertainties.  Since the photocurrents dominate the dark currents at all fluences, the dark currents do not affect our  measurements of relative responsivity.

The measured dark current after irradiation did not change in such a way to affect our ability to measure changes in the relative responsivity as a function of wavelength since our photocurrents are much greater than the measured dark currents after irradiation.

\subsection{Results for InGaAs Photodiodes}
\label{results}

We used the RERP facility to irradiate the InGaAs photodiodes in Table~\ref{tab:Photodiode} according to the exposure history in Table~\ref{tab:radruns}.  Before any radiation exposure, we scanned each photodiode and measured its dark current to determine $R_\lambda(0)$ in eq.(\ref{eqn:deltaNTNR}).  The dark currents from these baseline measurements are given in Table~\ref{tab:baselinedark}.  

\begin{figure}
\plotone{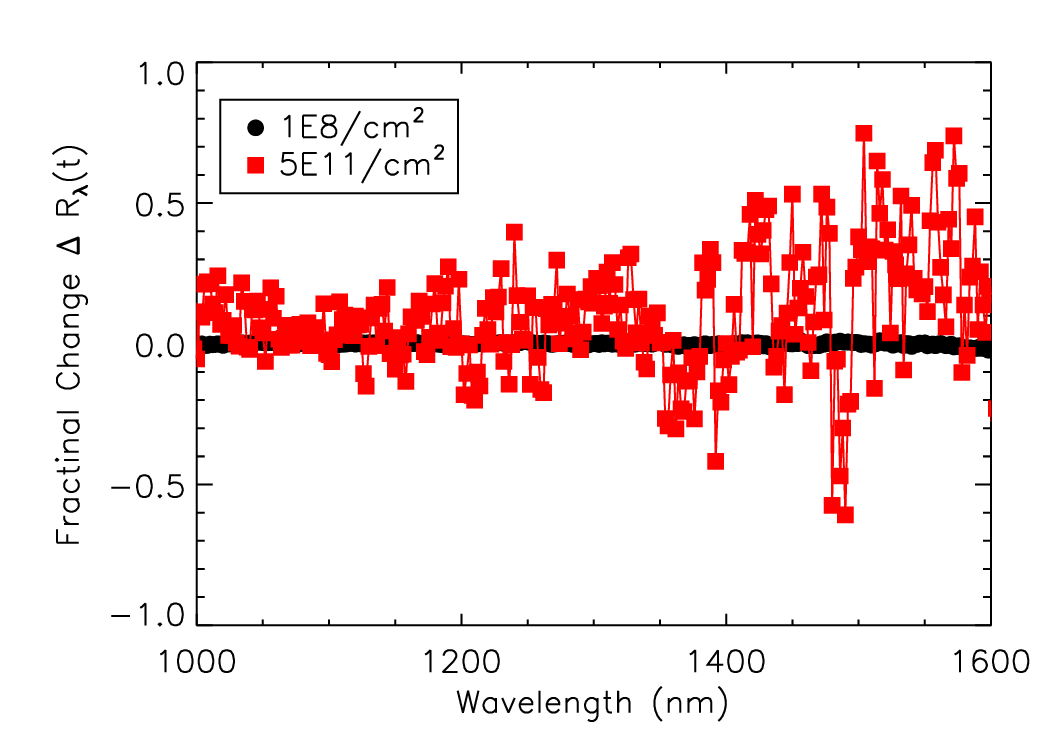}
\caption[]{$\Delta R_\lambda(t)$ for the PerkinElmer photodiode after exposure to 30 MeV protons at $1 \times 10^8$ and $5 \times 10^{11}$ protons/cm$^2$.  This behavior is representative of the behavior seen for all test InGaAs photodiodes at these fluences.}
\label{fig:PEfracChange}
\end{figure}

After each exposure we determined  $\Delta R_\lambda(t)$ by rescanning the photodiodes and remeasuring the dark current. 
Fig.~\ref{fig:PEfracChange} shows $\Delta R_\lambda(t)$ for the PerkinElmer photodiode after exposure to 30 MeV protons at ${\mathcal F} = 10^8$ and $5 \times 10^{11}$ protons/cm$^2$.  There appears to be little evidence for damage at $10^8$ protons/cm$^2$; there is, however, significantly more damage at $5 \times 10^{11}$ protons/cm$^2$.  The behavior seen at 30 MeV for this PerkinElmer photodiode is qualitatively similar to that seen in all the InGaAs photodiodes we investigated.

\begin{figure}
\begin{center}
\plotone{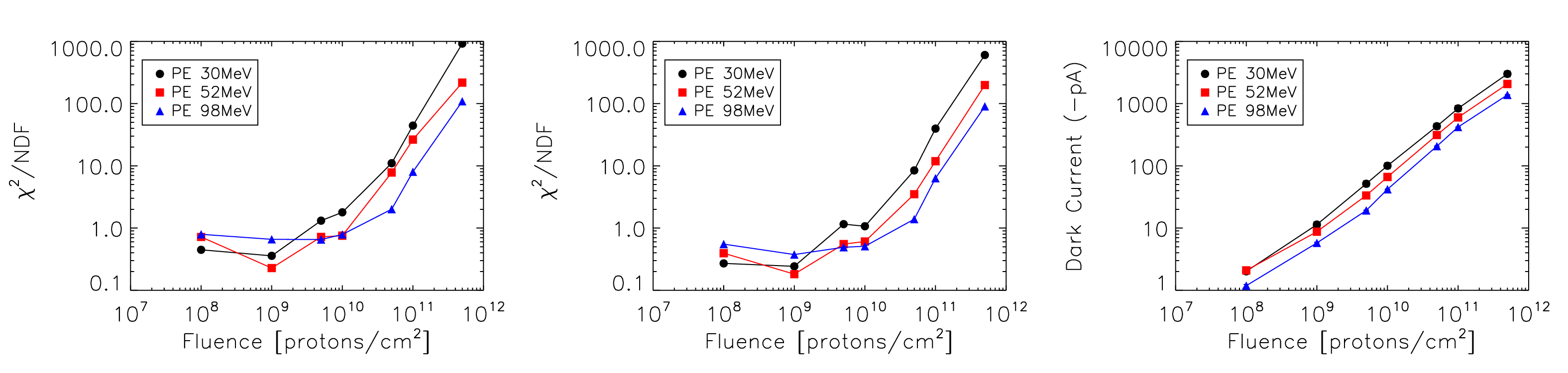}\\
\end{center}
\caption[]{Model fits to the radiation exposure data for the PerkinElmer InGaAs photodiodes. 
{\it left}: $\chi^2$/NDF for the $\mathcal{R}_\lambda = 0$ model as a function of fluence for the three proton irradiation energies.  Below $10^{10}$ protons/cm$^2$$\Delta R_\lambda(t)$ shows little evidence for damage; above $10^{10}$ protons/cm$^2$, the value of $\chi^2$/NDF increases significantly with fluence, which implies increasing damage to the photodiode with radiation exposure. 
{\it middle}: $\chi^2$/NDF for the wavelength-dependent $\mathcal{R}_\lambda$ model, where $\Delta R_\lambda(t)$ is fit with a fifth order polynomial.  The wavelength-dependent model clearly does not account for the increasing radiation damage above $10^{10}$ protons/cm$^2$.   
{\it right}:  The (negative) dark current as a function of fluence for the three proton energies.  The increases in dark current at fluences greater than $10^9$ protons/cm$^2$ are well above the 12\% systematic error in the measurements.  The correlation between the rise in $\chi^2$/NDF above $10^{10}$ protons/cm$^2$ and the comparable rise in wavelength-independent dark current is apparent.  This correlation would account for the increase in $\chi^2$/NDF.
}
\label{fig:PEdamage}
\end{figure}

We analyzed the $\Delta R_\lambda(t)$ data for the InGaAs photodiodes with eq.(\ref{eqn:chisq}) as was done for the OSI Si photodiodes. 
We first describe in detail the results from the PerkinElmer photodiodes.  Since the behaviors of the remaining InGaAs photodiodes are  similar, we give less detailed descriptions for them.  
Fig.~\ref{fig:PEdamage} shows the results for the PerkinElmer photodiodes.
The left panel of Fig~\ref{fig:PEdamage} shows $\chi^2$/NDF for the $\mathcal{R}_\lambda = 0$ model as a function of fluence for the three proton irradiation energies.  $\Delta R_\lambda(t)$ shows little evidence for damage below ${\mathcal F} = 10^{10}$ protons/cm$^2$; above ${\mathcal F} = 10^{10}$ protons/cm$^2$, however, the value of $\chi^2$/NDF increases significantly with fluence, a result that implies increasing damage to the photodiode with radiation exposure.  The middle panel shows $\chi^2$/NDF for the wavelength-dependent $\mathcal{R}_\lambda$ model, where $\Delta R_\lambda(t)$ is fit with a fifth order polynomial.  The wavelength-dependent model clearly does not account for the increasing radiation damage above $10^{10}$ protons/cm$^2$, which implies that the degradation in $\Delta R_\lambda(t)$ is due to wavelength-independent effects.  The right panel shows the (negative) dark current as a function of fluence for the three proton energies.  The increases in dark current are well above the 12\% systematic error in the measurements.  The correlation between the rise in $\chi^2$/NDF of  $\sim 3$ orders of magnitude above $10^{10}$ protons/cm$^2$ and the comparable rise in wavelength-independent dark current is apparent.  The $\chi^2$/NDF behavior in the  middle panel can be qualitatively understood as follows.  $I_\lambda$ is relatively insensitive to radiation damage at the fluences investigated here.  At low fluence, $I_\lambda(t)$ dominates $D(t)$ and $\Delta R_\lambda(t)$ shows little evidence of radiation damage.   As $D(t)$ increases with fluence, $D$ grows to dominate $I_\lambda$ and dark current adds increasingly large fluctuations to each monochromator wavelength measurement.  These fluctuations are well in excess of $\sigma$.  Since the sensitivity of COTS InGaAs photodiodes falls at longer wavelengths, while magnitude of the fluctuations in dark current are independent of wavelength, the relative importance of the fluctuations in $\Delta R_\lambda(t)$ increases at longer wavelengths.  This behavior in apparent in Fig.~\ref{fig:PEfracChange}.

\begin{figure}
\begin{center}
\plotone{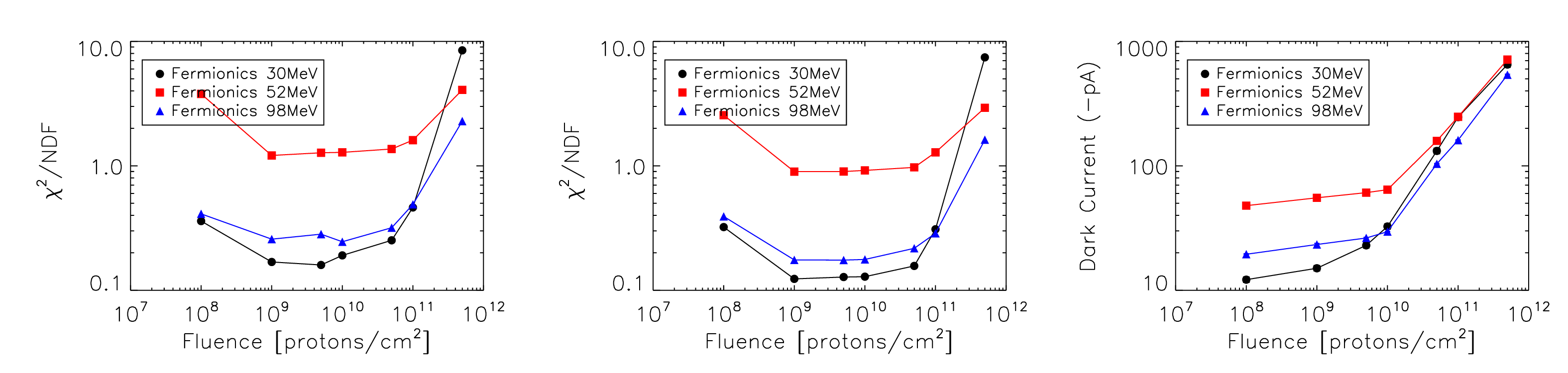}\\
\end{center}
\caption[]{Model fits to the radiation exposure data for the Fermionics InGaAs photodiodes; see Fig.~\ref{fig:PEdamage}.  
{\it left}: $\chi^2$/NDF for the $\mathcal{R}_\lambda = 0$ model as a function of fluence for the three proton irradiation energies.  
{\it middle}: $\chi^2$/NDF for the wavelength-dependent $\mathcal{R}_\lambda$ model.
{\it right}:  The (negative) dark current as a function of fluence for the three proton energies.}
\label{fig:Fermidamage}
\end{figure}

\begin{figure}
\begin{center}
\plotone{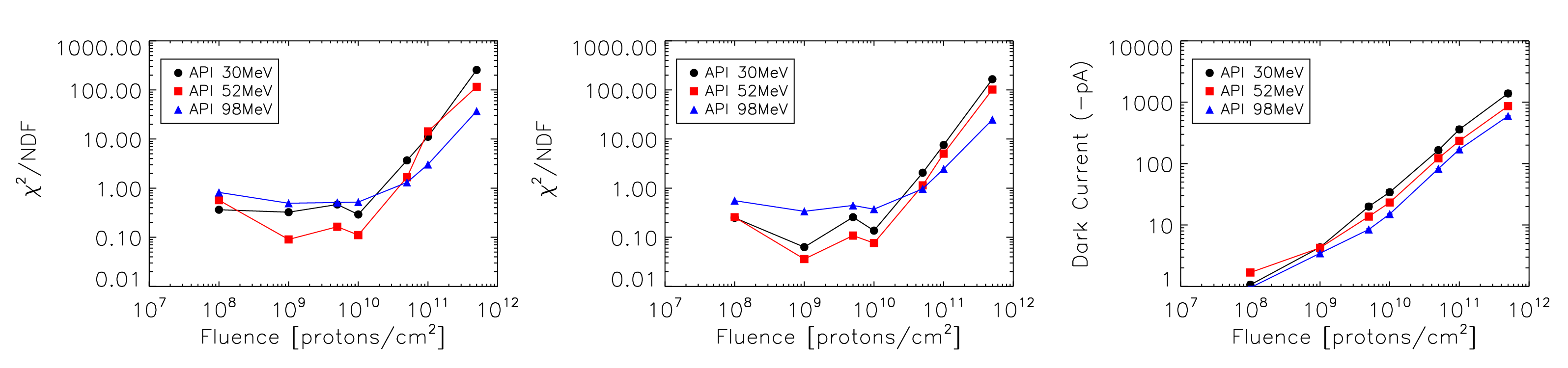}\\
\end{center}
\caption[]{
Model fits to the radiation exposure data for the API InGaAs photodiodes; see Fig.~\ref{fig:PEdamage}.  
{\it left}: $\chi^2$/NDF for the $\mathcal{R}_\lambda = 0$ model as a function of fluence for the three proton irradiation energies.  
{\it middle}: $\chi^2$/NDF for the wavelength-dependent $\mathcal{R}_\lambda$ model.
{\it right}:  The (negative) dark current as a function of fluence for the three proton energies.}
\label{fig:APIdamage}
\end{figure}

\begin{figure}
\begin{center}
\plotone{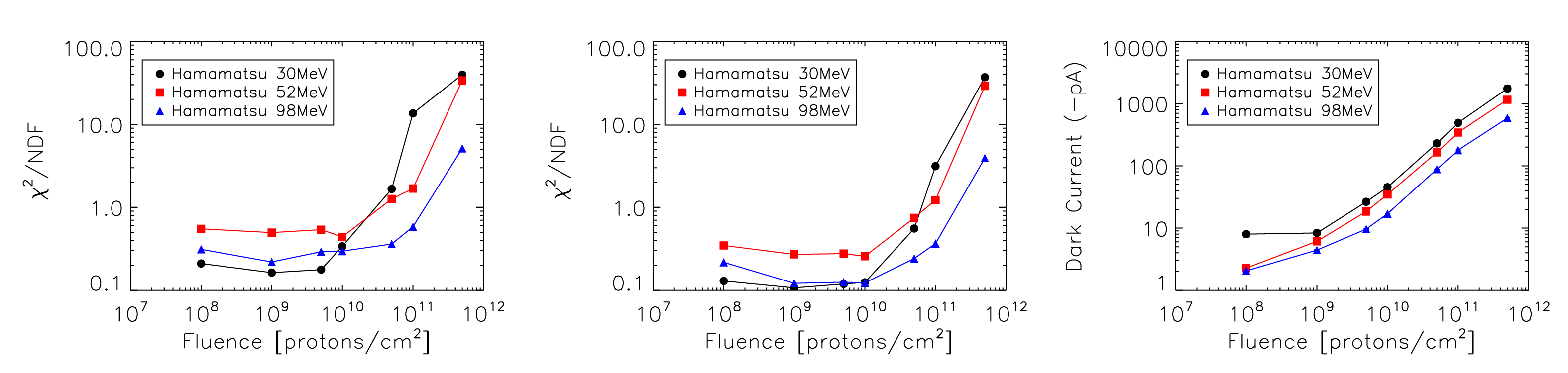}\\
\end{center}
\caption[]{Model fits to the radiation exposure data for the Hamamatsu InGaAs photodiodes; see Fig.~\ref{fig:PEdamage}.  
{\it left}: $\chi^2$/NDF for the $\mathcal{R}_\lambda = 0$ model as a function of fluence for the three proton irradiation energies.  
{\it middle}: $\chi^2$/NDF for the wavelength-dependent $\mathcal{R}_\lambda$ model.
{\it right}:  The (negative) dark current as a function of fluence for the three proton energies.}
\label{fig:Hamamatsudamage}
\end{figure}

\begin{figure}
\begin{center}
\plotone{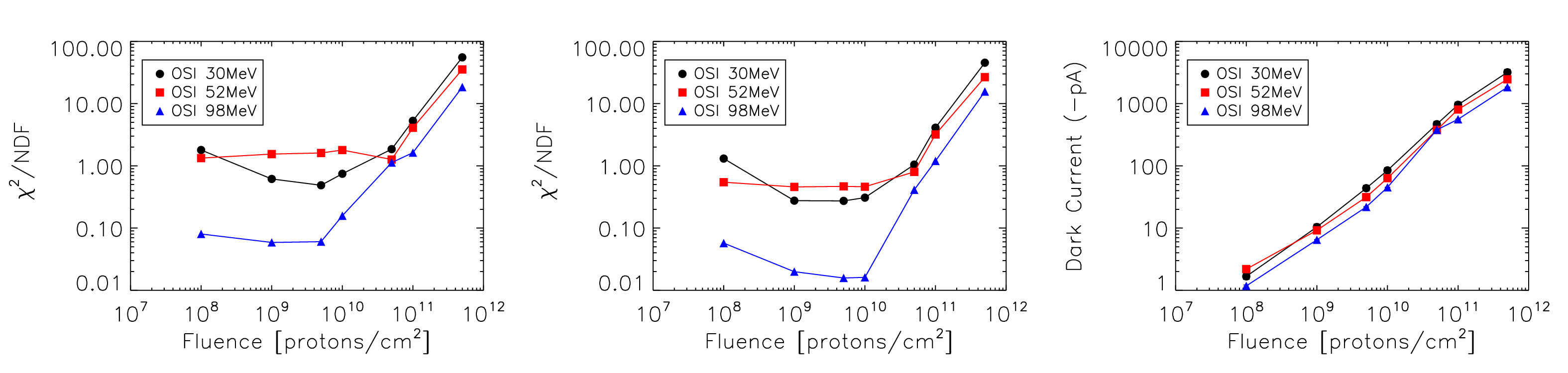}\\
\end{center}
\caption[]{Model fits to the radiation exposure data for the OSI InGaAs photodiodes; see Fig.~\ref{fig:PEdamage}.  
{\it left}: $\chi^2$/NDF for the $\mathcal{R}_\lambda = 0$ model as a function of fluence for the three proton irradiation energies.  
{\it middle}: $\chi^2$/NDF for the wavelength-dependent $\mathcal{R}_\lambda$ model.
{\it right}:  The (negative) dark current as a function of fluence for the three proton energies.}
\label{fig:OSIdamage}
\end{figure}

The results for the remaining photodiodes analyzed with eq.(\ref{eqn:chisq}) are shown in Figs.~\ref{fig:Fermidamage} -- \ref{fig:OSIdamage}.  The behavior seen in  Fig.~\ref{fig:PEdamage} is qualitatively similar to what is seen for these photodiodes: radiation damage leads to significant increases in wavelength-independent dark current that affect photodiode response.  
More specifically, the PerkinElmer and OSI photodiodes show the greatest increases in dark current.  The API and Hamamatsu photodiodes show smaller increases.  The Fermionics photodiodes show the smallest relative changes.   However, the baseline dark currents for the Fermionics photodiodes are much greater than the others. 
The increase in dark current as a function of radiation exposure has also been reported for InGaAs Avalanche Photodiodes \citep{bib:Becker-Johnson-darkcurrent-deg}.  

\section{Absolute Response of InGaAs Photodiodes through 10 nm Narrow Band Filters}
\label{sec:absspecresponce}

Since our spectral response measurements for InGaAs photodiodes only determine relative changes in response as a function of wavelength, these measurements would not detect radiation damage that affects the overall photodiode response in a wavelength independent way -- a so-called ``graying'' of the response.  To evaluate whether graying has occurred, we developed a second apparatus, the Absolute Responsivity Apparatus (ARA), that measures absolute changes in the responsivity of the InGaAs photodiodes with a filament lamp viewed through a set of narrowband interference filters.

\subsection{Absolute Responsivity Apparatus}
\label{ARA}

The ARA consists of a 50 W QTH lamp, an automated shutter, a filter wheel containing seven narrowband, 1.25'' diameter Oriel interference filters, and hardware to mount the photodiode under test.  The QTH lamp is positioned 0.25~m from the photodiode and is powered by a Newport constant-current radiometric power supply.  We used the automated shutter to monitor the dark current before and after each filter measurement.  The seven narrowband filters in the filter wheel are given in Table \ref{tab:filters}.  
The photodiodes were read out 100 times per filter with a Keithley dual-channel picoammeter at a sample rate of 6~Hz to reduce read noise.  The 100 measurements were averaged to determine the photo-current.  

The ARA yields photo-currents that are $10^4 -10^5$ times greater than the dark current.  This large signal dominates the baseline dark currents, as well as variations in the read-out electronics that can add uncertainty to the measurements, so that the ARA is testing the overall photodiode response.

\subsection{Experiment Design}

Since the filters are 1.25'' in diameter, only one photodiode at a time can be measured in the ARA.  We therefore mounted the test and reference photodiodes on a sliding stage so that they could be moved in and out of the apparatus.  The measurements were carried out with the following cadence: the reference photodiode in all filters, the test photodiode in all filters, and then the reference photodiode in all filters once again.  We averaged the two reference photodiode measurements in each filter that bracket the test photodiode measurements.  Again we used the ratio of the response of the test photodiode to the average response of the reference photodiode to look for damage due to radiation exposure.

In analogy with eq.(\ref{eqn:NTNR}), we define the absolute spectral response 
of the test photodiode relative to the NIST calibrated photodiode in bandpass $\Delta \lambda$ at time $t$.  $A_{\Delta \lambda}(t)$, is given by 
\begin{equation}
\label{eqn:TNR}
A_{\Delta \lambda}(t) = \left[\frac{I_{\Delta \lambda}(t)}{<N_{\Delta \lambda}>}\right],
\end{equation}
where $I_{\Delta \lambda}(t)$ are measurements of the response of the test photodiode,  $<N_{\Delta \lambda}>$ is the average of the reference photodiode measurements flanking the test photodiode, and $\Delta \lambda$ is the filter bandpasses given in Table~\ref{tab:filters}.  The fractional change in the absolute spectral response is then 
\begin{equation}
\label{eqn:fracTNR}
\Delta A_{\Delta \lambda}(t) = A_{\Delta \lambda}(t)/A_{\Delta \lambda}(0) - 1.
\end{equation}
In our narrowband filter experiments we compared photodiode measurements after six different exposures -- $1\times 10^9$ protons/cm$^2$, $5\times 10^9$ protons/cm$^2$, $\ldots$, $5\times 10^{11}$ protons/cm$^2$ and we used the $1\times 10^9$ protons/cm$^2$ exposure as the $A_{\Delta \lambda}(0)$.  The narrowband measurement program was initiated after the $1\times 10^9$ protons/cm$^2$ exposure.  

\subsection{Systematic Uncertainties}

We determined the systematic uncertainties introduced by the filter-testing apparatus by once again measuring the unexposed control PerkinElmer InGaAs photodiode.  We made five back-to-back measurements of this photodiode (without power cycling the apparatus) through the filters listed in Table~\ref{tab:filters} on seven separate days from October 2009 to May 2011.  We combined the control photodiode measurements on separate days to establish the response expected for a photodiode that did not change as a result of radiation exposure.   We then compared this baseline behavior with the irradiated photodiodes as a test of the null hypothesis -- is the irradiated photodiode consistent with the hypothesis that it did {\it not} gray?  

To model the response of a InGaAs photodiode that does not gray, we constructed a set of simulated experiments in each filter with the data from the unexposed control PerkinElmer photodiode.  For these simulations, we first created a histogram of all possible $\Delta A_{\Delta \lambda}$ values that pair a single measurement on one day with a single measurement on any other day (e.g., measurement \#1 on day 1 with measurement \#4 on day 3).  We did not pair measurements on the same day because they were made without power-cycling the apparatus, unlike the measurements of the test photodiodes made after each exposure.  

For any given filter there are 525 $\Delta A_{\Delta \lambda}$ entries in the histogram.  We then constructed simulated experiments by 
drawing five values of $\Delta A_{\Delta \lambda}$ from this histogram and computing their mean and standard deviation.  This procedure 
simulates the statistics we computed through each filter for each photodiode after the radiation exposure program was completed.   (Using the first exposure as the reference, there are five $\Delta A_{\Delta \lambda}$ values from the six exposures.)  In all we constructed 10,000 experiments through each filter for a total of 70,000 simulated experiments.   

\begin{figure}
\plotone{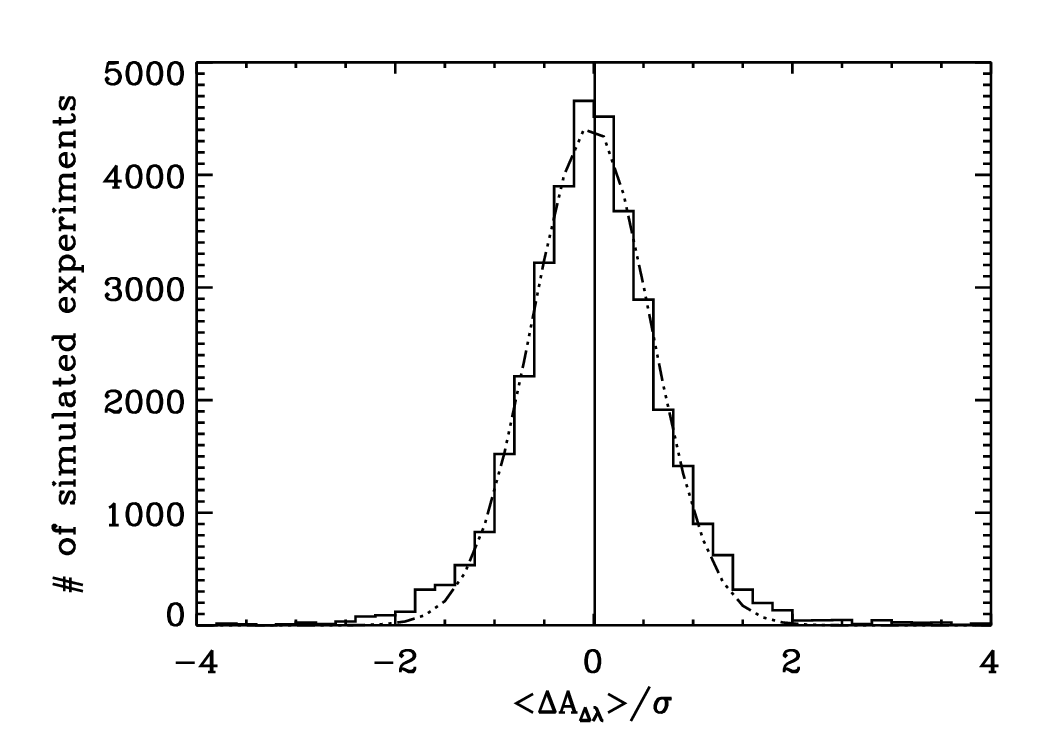} 
\caption[]{The behavior expected for a photodiode whose bandpass response does not gray.  Plotted are the number of standard deviations that the mean of five measurements of $\Delta A_{\Delta \lambda}$ is displaced from zero, $<\Delta A_{\Delta \lambda}>/\sigma$, for 70,000 simulated experiments.  Superposed is a Gaussian with mean -0.05 and standard deviation 0.6.}
\label{fig:simuatedExperiments}
\end{figure}

In Fig.~\ref{fig:simuatedExperiments}, we plot the results of the simulated experiments.  This figure histograms the number of standard deviations that the mean is displaced from zero, $<\Delta A_{\Delta \lambda}>/\sigma$, for all 70,000 simulated experiments.  Superposed on this distribution is a Gaussian, the distribution expected for our simulated experiments if they differed from one another as a result of measurement error.  The mean of this distribution falls close to zero, as expected in this case.  The width is narrower than 1 because we are averaging five values that are all drawn from a normal distribution.  

Fig.~\ref{fig:simuatedExperiments} shows the behavior expected from a photodiode whose response has suffered no graying damage.

\subsection{Results for InGaAs Photodiodes}
\begin{figure}
\plotone{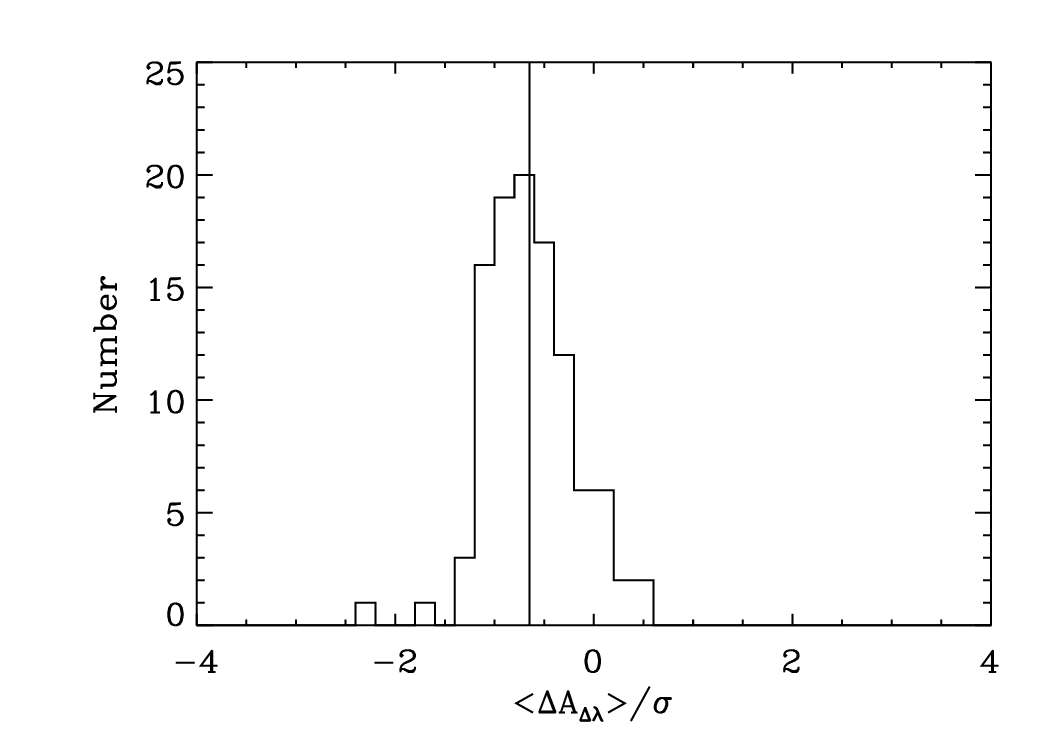} 
\caption[]{The behavior of InGaAs test photodiodes.  Plotted are the number of standard deviations that the mean of five measurements of $\Delta A_{\Delta \lambda}$ is displaced from zero, $<\Delta A_{\Delta \lambda}>/\sigma$, for the total data set.  The shape and width of this distribution in Fig.~\ref{fig:realExperiments} is consistent with the simulated experiments Fig.~\ref{fig:simuatedExperiments}.  The offset of the peak from zero likely results from making an explicit choice for $A_{\Delta \lambda}(0)$ in $\Delta A_{\Delta \lambda}(t)$, as described in the text.
}
\label{fig:realExperiments}
\end{figure}

We measured the absolute spectral response of the InGaAs photodiodes in Table~\ref{tab:Photodiode} using the ARA.  Since we initiated this program after the photodiodes had been exposed to a fluence of ${\mathcal F} = 10^9$ protons/cm$^2$, we used the measurements at $10^9$ protons/cm$^2$ as the baseline $A_{\Delta \lambda}(0)$ in eq.(\ref{eqn:fracTNR}).  We then remeasured the test photodiodes after the remaining exposures in Table~\ref{tab:radruns}.  The results are shown in Fig.~\ref{fig:realExperiments}.  This figure is to be compared with Fig.~\ref{fig:simuatedExperiments}, which shows the behavior expected from a photodiode whose response has suffered no graying.  The shape and width of the data distribution in Fig.~\ref{fig:realExperiments} is reasonably consistent with Fig.~\ref{fig:simuatedExperiments}.  The peak, however, is offset from zero.  We infer this offset  is the result of making an explicit choice for $A_{\Delta \lambda}(0)$ in $\Delta A_{\Delta \lambda}(t)$.  In the simulated experiments, all days were treated equally in computing $A_{\Delta \lambda}(0)$ in eq.(\ref{eqn:TNR}), a procedure that mitigates the bias introduced by an explicit choice for $A_{\Delta \lambda}(0)$.  We tested this hypothesis by choosing a different day for $A_{\Delta \lambda}(0)$ in evaluating eq.(\ref{eqn:fracTNR}) and we found that the peak moved significantly, consistent with the hypothesis.

We infer from the results shown in Fig.~\ref{fig:realExperiments} that there was no graying of the InGaAs photodiode response at the radiation doses investigated here.  This suggests there would be negligible damage affecting the absolute spectral response of InGaAs photodiodes with the radiation exposure expected at L2.

\section{Summary}
In this investigation, we report on radiation hardness studies of several COTS InGaAs photodiodes and one COTS Si photodiode exposed to ionizing protons with energies of 30, 52, and 98 MeV, at fluences up to  ${\mathcal F} =  5 \times 10^{11}$~cm$^{-2}$ at the RERP at the IUCF.  

We scanned the relative spectral response and measured the dark current with the IRRMA apparatus we developed for this investigation.  We found that both the Si and InGaAs photodiodes experience radiation damage as the fluence increases.  The Si photodiodes showed wavelength-dependent radiation damage, particularly at wavelengths longer than $\lambda = 700$ nm, primarily as a result of damage to their responsivity.  
The InGaAs photodiodes, however, showed evidence for wavelength-independent damage as a result of significant increases in their dark current as the fluence increased.  

We used the ARA apparatus we developed to measure absolute changes in the responsivity of the InGaAs photodiodes.   This investigation was designed to determine whether there was an overall graying of the response of the InGaAs photodiodes after radiation exposure.  By comparing the ARA measurements to simulated experiments constructed from baseline measurements of the InGaAs photodiodes before radiation exposure, we found that the test photodiodes showed little evidence for graying of their response.

\acknowledgments
\section*{Acknowledgments}
This work was supported by the Office of Science of the US DOE.  We would like to thank Barbara Von Przewoski at IUCF RERP for her help and Frederic Laforce at PerkinElmer for kindly supplying the PerkinElmer photodiodes used in this investigation.  We gratefully acknowledge the support of grants 6706131 and 6890462 from Lawrence Berkeley National Laboratories to Indiana University in this research.

\bibliographystyle{natbib}

%
%

\begin{deluxetable}{ccccc}
\tablecaption{Photodiodes tested.}

\tablewidth{0pt}
\tablehead{\colhead{Manufacturer}&\colhead{Semiconductor} & \colhead{Active Area} & \colhead{Package} & \colhead{Part}  \\\colhead{} & \colhead{Type}& \colhead{Diameter}& \colhead{Type}& \colhead{Number}  }
\startdata
Advanced Photonix Inc & InGaAs& 1.5mm& TO-39& SD 060-11-41-211  \\
Fermionics  & InGaAs& 2mm& TO-5& FD2000W   \\
Hamamatsu  & InGaAs& 2mm& TO-5& G8370-82  \\
OSI Optoelectronics & InGaAs& 3mm& TO-5& FCI-InGaAS-3000   \\
OSI Optoelectronics & Si& 3mm& TO-5& OSD15-0   \\
PerkinElmer  & InGaAs& 2mm& TO-5& C30642GH   \\
\enddata
\label{tab:Photodiode}
\end{deluxetable}

\begin{deluxetable}{ccccc}
\tablecaption{NIST calibrated reference photodiodes.}

\tablewidth{0pt}
\tablehead{\colhead{Manufacturer}&\colhead{Semiconductor} & \colhead{Active Area} & \colhead{Package} & \colhead{Part}  \\\colhead{} & \colhead{Type}& \colhead{Diameter}& \colhead{Type}& \colhead{Number}  }
\startdata
Hamamatsu  & InGaAs& 3mm& TO-8& G5851-23\\
Hamamatsu  & Si& 3.6mm& TO-5& S1336-44BK\\
\enddata
\label{tab:Photodiodes-nist}
\end{deluxetable}

\begin{deluxetable}{ccc}
\tabletypesize{\small}
\tablecaption{RERP exposure history at 30, 52, and 98 MeV.}
\tablewidth{0pt}
\tablehead{ \colhead{Date} & \colhead{Cumulative Fluence} & \colhead{Beam}\\ \colhead{} & \colhead{[protons/cm$^2$]} & \colhead{Configuration}}
\startdata
5/15/09 &  $1\times10^8$&Narrow \\
6/06/09&   $5\times10^8$&Narrow \\
7/31/09 & $1\times10^9$ &Narrow\\
11/11/09 & $5\times10^9$ &Narrow\\
2/22/10 &  $1\times10^{10}$&Wide \\
6/18/10 &  $5\times10^{10}$&Wide \\
8/09/10 &  $1\times10^{11}$&Wide \\
9/24/10 &  $5\times10^{11}$&Wide \\
\enddata
\label{tab:radruns}
\end{deluxetable}

\begin{deluxetable}{cc}
\tabletypesize{\small}
\tablecaption{Value for $\sigma$ used in $\chi^2$ calculation for each photodiode manufacturer determined from the baseline measurements of each of the photodiodes relative to the other.}
\tablewidth{0pt}
\tablehead{\colhead{Photodiode} & \colhead{$\sigma$} }
\startdata
API        & 0.009 \\
Fermionics & 0.02  \\
Hamamatsu  & 0.01  \\
OSI        & 0.01  \\
OSI Si     & 0.002 \\
PerkinElmer& 0.008 \\
\enddata
\label{tab:chisqvalues}
\end{deluxetable}

\begin{deluxetable}{cccc}
\tabletypesize{\small}
\tablecaption{Dark current measurements for InGaAs photodiodes prior to irradiation; a different set of photodiodes was used for the 3 program irradiation energies.}
\tablewidth{0pt}
\tablehead{\colhead{Photodiode} & \colhead{30 MeV} & \colhead{52 MeV} & \colhead{98 MeV}\\\colhead{Manufacturer} & \colhead{Dark Current  [pA]}& \colhead{Dark Current  [pA]} & \colhead{Dark Current  [pA]}}
\startdata
API & -0.98       & -1.45       & -1.53\\
Fermionics & -11.55&  -45.81& -18.72\\
Hamamatsu  & -3.02 & -3.20 & -2.75\\
OSI  & -1.93       & -2.50       & -2.89\\
PerkinElmer & -1.73        & -2.83      & -1.92\\
\enddata
\label{tab:baselinedark}
\end{deluxetable}

\begin{deluxetable}{cc}
\tabletypesize{\small}
\tablecaption{Dark current measurements for Si photodiodes prior to irradiation; a different photodiode was exposed to 30 MeV protons at the fluences given in the first column.}
\tablewidth{0pt}
\tablehead{\colhead{Fluence} & \colhead{Dark Current}\\\colhead{[protons/cm$^2$]} & \colhead{[pA]} }
\startdata
$5\times10^8$&-0.073\\
$1\times10^9$&-0.065\\
$5\times10^9$&-0.064\\
$1\times10^{10}$&-0.11\\
$5\times10^{10}$&-0.076\\
$1\times10^{11}$&-0.084\\
$5\times10^{11}$&-0.078\\
\enddata
\label{tab:baselinedarksi}
\end{deluxetable}

\clearpage

\begin{deluxetable}{ccc}
\tabletypesize{\small}
\tablecaption{Narrowband filters in the filter-testing apparatus.}
\tablewidth{0pt}
\tablehead{\colhead{$\lambda_{\textrm{center}}$} & \colhead{FWHM} & \colhead{$T_\textrm{max}$} \\
 \colhead{(nm)} & \colhead{(nm)} & \colhead{\%}\\}
\startdata
 700 & 10 & 50\\
 800 &  10 & 45\\
 900 &  10 & 45\\
 1000 &  10 & 40\\
 1050 &  10 & 40\\
 1200 &  10 & 35\\
 1550 &  10 & 30\\
\enddata
\label{tab:filters}
\end{deluxetable}

\begin{deluxetable}{cc}
\tabletypesize{\small}
\tablecaption{Systematic uncertainties in $\Delta A_{\Delta\lambda}(t)$ for narrowband filter measurements.}
\tablewidth{0pt}
\tablehead{\colhead{$\lambda_{\textrm{center}}$} &\colhead{Simulated}\\
\colhead{(nm)} & \colhead{Mean}}
\startdata
   700&    -0.0003 $\pm$ 0.004\\
   800&   0.002 $\pm$ 0.005\\
   900&   -0.001 $\pm$ 0.007\\
  1000&   -0.001 $\pm$ 0.004\\
  1050&  0.0004 $\pm$ 0.01\\
  1200&    0.009 $\pm$ 0.02\\
  1550&   -0.004 $\pm$ 0.008\\
\enddata
\label{tab:filter_sytematics}
\end{deluxetable}

\clearpage

\clearpage

\clearpage

\clearpage

\clearpage

\clearpage

\clearpage

\clearpage

\end{document}